\documentclass[aip,apl,twocolumn,amsmath,amssymb]{revtex4}
\usepackage{bm}
\usepackage{amssymb}
\usepackage{amsmath}
\usepackage{latexsym}
\usepackage[dvips]{graphicx}
\begin{document}
\author{Stanislav Stoupin}
\email{sstoupin@aps.anl.gov}
\affiliation{Advanced Photon Source, Argonne National Laboratory, Lemont, Illinois 60439, USA}
\author{Sergey Antipov}
\affiliation{Euclid Techlabs LLC, Solon, OH 44139, USA}

\author{Alexander V. Kolyadin}
\affiliation{New Diamond Technology LLC, St. Petersburg, 197706, Russia}

\author{Andrey Katrusha}
\affiliation{New Diamond Technology LLC, St. Petersburg, 197706, Russia}

\author{James E. Butler}
\affiliation{Euclid Techlabs LLC, Solon, OH 44139, USA}

\title{Large-surface-area diamond (111) crystal plates for applications in high-heat-load wavefront-preserving x-ray crystal optics\footnote{Submitted for publication to J. Synchrotron Rad.}}
\begin{abstract}
We report fabrication and results of high-resolution X-ray topography characterization of diamond single crystal plates with a large surface area 
(10~$\times$~10 mm$^2$) and (111) crystal surface orientation for applications in high-heat-load X-ray crystal optics. The plates were fabricated by laser cutting of the (111) facets of diamond crystals grown using high-pressure high-temperature method. The intrinsic crystal quality of a selected 3~$\times$~7~mm$^2$ crystal region of one of the studied samples was found to be suitable for applications in wavefront-preserving high-heat-load crystal optics. The wavefront characterization was performed using sequential X-ray diffraction topography in the pseudo plane wave configuration and data analysis using rocking curve topography. The variation of the rocking curve width and peak position measured with a spatial resolution of 13~$\times$~13~$\mu m^2$ over the selected region were found to be less than one microradian. 
\end{abstract}

\maketitle


\section{Introduction}

Diamond crystal plates with (111) surface orientation are of primary importance for high-heat-load X-ray monochromators (\cite{Grubel96,Fernandez97,Yabashi07}) at light source facilities which generate intense hard X-rays (wavelengths on the order of 1~$\rm \AA$). The choice of the surface orientation is due to the largest intrinsic energy bandwidth of the 111 Bragg reflection ($\Delta E/E \simeq 5.7 \times 10^{-5}$) and the resulting greater flux of the reflected (monochromatized) X-rays as compared with those of higher-order reflections. Although these quantities are greater for Ge 111 ($\Delta E/E \simeq 3.1 \times 10^{-4}$) or  Si  111 ($\Delta E/E \simeq 1.3 \times 10^{-4}$), commonly used crystals/reflections in the high-heat-load monochromators diamond as a monochromator crystal has two main advantages. 
One of them is the reduction in operating costs related to the choice of thermal management solution. Water cooling can be used for diamond due to its exceptional thermal properties while liquid nitrogen cooling is often required for Si and Ge to minimize heat-load-induced crystal deformations. 
The second advantage is the possibility to split the incident intense X-ray beam into the transmitted and reflected branches, an approach pioneered at the TROIKA beamline at the ESRF (European Synchrotron Radiation Facility, Grenoble, France)~\cite{ANielsen94,Grubel94,Grubel96}. This approach helps to improve productivity of a beamline by conducting two experiments simultaneously. Such X-ray-optics-based beam-multiplexing approach gains even more importance for hard X-ray free-electron lasers (XFELs) where the operating cost per experiment is considerably greater compared to that of a storage-ring-based synchrotron sources. Diamond (111) beam-multiplexing optics for XFELs has been recently demonstrated \cite{Stoupin14,Zhu13}. It has become apparent that detailed X-ray diffraction characterization of diamond (111) optics is a necessary step which reveals details about distortion of the crystal lattice and its influence on preservation of the radiation wavefront. 

Preserving the wavefront of intense X-ray beams is a major challenge for modern X-ray optics as coherence properties of X-ray sources are rapidly improving and the experiments become more demanding for the beam quality. Besides the effects of intrinsic distortion, a heat-load-induced crystal distortion is an issue for certain experiments at third generation synchrotrons due to the high repetition rate (a few MHz) as compared to the presently operational low repetition rate XFELs (up to 120 Hz). Despite a much greater instantaneous flux of XFEL (i.e., number of photons per pulse) this dramatic difference in the repetition rate yields $\approx$~100X  greater average heat load on double-crystal monochromators at synchrotrons. Furthermore, next-generation synchrotrons and high-repetition rate XFELs will likely generate even greater average flux densities. Thus, a next  generation of X-ray optics is required to complement the rapid progress in X-ray sources, and in particular for hard x~rays, wavefront preserving high-heat-load diffracting crystal optics is much desired. Although the advantages of using diamond for the next generation of high-heat-load crystal optics are clear, the major obstacle is the small size of the available high-quality single crystals, and, as a result, deterioration of monochromator performance due to difficulties in thermal management (e.g, insufficient coupling to a crystal holder due to small available surface area). 

The typical size of (111) crystals in high-heat-load monochromators is about 4~$\times$~8~mm$^2$ (e.g., \cite{Yabashi07}) and the typical working area (region of high crystal quality) is approximately 2~$\times$~5~mm$^2$ for selected crystals (e.g., \cite{Stoupin14}). Such sizes are related to the crystal plate fabrication procedure where the (111) plate is sliced across the (001) growth sector which typically has the best crystal quality \cite{Burns09,Sumiya12,Stoupin_DRM14}. As a result the high quality crystal region is adjacent to one of the edges of the plate \cite{Stoupin14, Stoupin_DRM14}. X-ray topography studies of larger (8~$\times$~8~mm$^2$)  diamond plates fabricated entirely from the (111)-growth-sector in crystals grown in the [111] direction were performed, however regions of high crystal quality were found to be too small for practical applications \cite{Goto12}.

In this work we report production of 1-mm-thick diamond (111) crystal plates with large total surface area of about 10~$\times$~10~mm$^2$.  The plates were extracted from the (111) growth sectors of a large type IIa diamond crystal grown in the [001] direction using high-pressure high-temperature method by New Diamond Technologies LLC \cite{Deljanin15}. The crystal plates were characterized with Laue X-ray topography using polychromatic (white-beam) synchrotron radiation and with rocking curve topography in the pseudo plane wave configuration. The performed X-ray characterization revealed a nearly defect-free crystal region having a very large size of 3~$\times$~7~mm$^2$ for a selected plate. The region was nearly centered on the plate, which is suitable for the traditional crystal mounting method using large regions adjacent to crystal edges for thermal contact (e.g., \cite{Fernandez97,Schildkamp02}). The variation of the width and peak position of the rocking curves (with  13~$\times$~13~$\mu m^2$ spatial resolution) across the region was found to be less than one microradian. Our findings suggest that application of the studied selected crystal plate in high-heat-load monochromators can substantially improve their performance parameters and has the potential to achieve preservation of the radiation wavefront to less-than-one-microradian level under high-heat-load conditions. 

\section{Crystal fabrication}
\begin{figure}
\centering\includegraphics[width=0.48\textwidth]{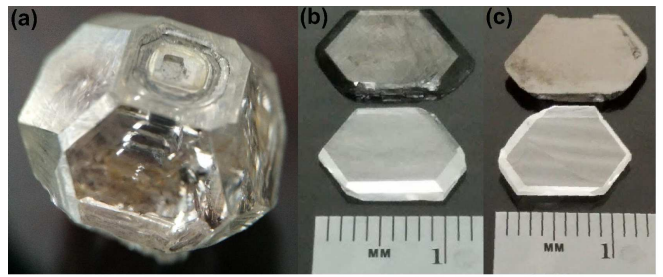}
\caption{(a) The semi-cut HPHT diamond from which the (111) plates were fabricated.  The approximate dimensions of the crystal are 18 mm across. 
(b) A photograph of the two plates from the as-grown side (plate A on the top and plate B the bottom). 
(c) A photograph of the two plates from the laser sliced surface (plate A on the top and plate B on the bottom). 
Slightly misaligned facets are observed on the polished side of plate B.}
\label{fig:dia}
\end{figure} 

A very large (62 ct) single crystal diamond was grown in the [001] direction by a proprietary high-pressure high-temperature (HPHT) process by New Diamond Technology (Russia).
The diamond was type IIa (no detectable nitrogen impurities by FTIR). The as-grown diamond displayed several large (111) faces.
The bottom part of the as-grown crystal containing the diamond seed was cut off. The resulting (semi-cut) diamond crystal with weight of 42.76 ct is shown in Fig.~\ref{fig:dia}(a). Plates with thicknesses of about 1~mm containing major (111) surfaces were laser sliced from the semi-cut diamond. 
The laser slicing was perfomed approximately parallel to the (111) faces with precision of $\approx$~1~deg. 
The photographs of the plates are shown in Fig.~\ref{fig:dia}(b,c). On one of the plates, the laser sliced surface was left as cut (plate A), while for the other, the laser cut surface was polished (plate B) using gemstone polishing equipment. A few slightly misaligned facets were observed on the polished side. 
Polishing of diamond surfaces has an extremely anisotropic wear rate with traditional iron skive polishing and the 111 surface surface is nearly impossible to polish (see \cite{Bouwelen99} and references therein). As a result, it is not surprising to observe slightly misaligned facets on the laser cut surface from the polishing attempts.

\section{White beam X-ray topography in the Laue geometry}
\begin{figure}
\centering\includegraphics[width=0.48\textwidth]{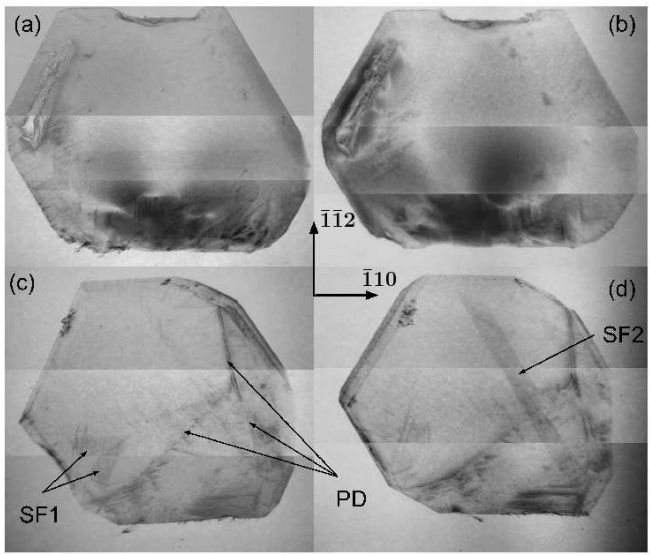}
\caption{White beam X-ray Laue topographs of the diamond crystal plates: (a) plate A $\bar{1}11$ reflection, (b) plate A $1\bar{1}1$ reflection , (c) plate B $\bar{1}11$ reflection, and (d)  plate B $1\bar{1}1$ reflection.  For plate B locations of stacking faults (SF1 and SF2) and plastic deformation (PD) are indicated with arrows. The horizontal lines dividing the topographs into 3 regions with slight variation in the background contrast are due to stitching of images from different X-ray films used for each crystal (due to the limited vertical size of the X-ray beam). The reference directions in the reciprocal lattice (common for all topographs) are shown in the center of the figure.}
\label{fig:wbxt}
\end{figure} 

Initial screening of crystal quality was performed using white beam topography in the Laue geometry. The experiment was performed at 1-BM beamline of the Advanced Photon Source (Argonne National Laboratory, IL, USA). In white (polychromatic) beam X-ray Laue topography footprints of multiple diffracted beams (topographs) are registered simultaneously on an X-ray film placed behind a studied single crystal. Various defects of crystal lattice are revealed such as inclusions, stacking faults, dislocations and plastic deformation (surface defects). Defects can remain invisible on certain topographs and reveal themselves on others. This is true even for topographs corresponding to different reflections of the same family having the same structure factor since defects do not obey the crystal symmetry rules. 
White beam X-ray topographs corresponding to Laue reflections $\bar{1}11$ and $1\bar{1}1$ are shown in Fig.~\ref{fig:wbxt} for each crystal plate. For plate A (Fig.~\ref{fig:wbxt}(a) and (b)), a strong kinematic diffraction contrast across the entire plate is seen due to the strain field propagating from the bottom edge of the plate. 
For plate B, two main types of defects are clearly observed: stacking faults and plastic deformation. In Fig.~\ref{fig:wbxt}(c) corresponding to $\bar{1}11$ reflection a double V-shaped stacking fault (SF1) is found close to one of the crystal edges. In Fig.~\ref{fig:wbxt}(d) corresponding to $1\bar{1}1$ reflection another stacking fault (SF2) is clearly observed. It is noted that SF1 is not clearly seen on topograph (c), and, similarly diffraction contrast due to SF2 is weak on topograph (d), which illustrates the complementary character of the topographs. On both topographs of plate B, signs of plastic deformation are observed. Lines marked PD on the topographs correspond to boundaries between visible facets on the crystal left after polishing. 
Nevertheless, a large nearly defect-free area is present which prompts further, more detailed X-ray study of this crystal plate.  

\section{Rocking curve X-ray topography in the pseudo plane wave Bragg geometry}
X-ray diffraction topography in pseudo plane wave configuration is a standard crystal characterization technique with high sensitivity to local crystal strain ($\approx 10^{-6}$). A number of applications of the technique in studies of diamond single crystals of different grades and orientation have been reported in the recent years \cite{Hoszowska_JPD01,Macrander05,Tamasaku05,Burns09,Goto12,Prokhorov13,Stoupin_DRM13,Stoupin14}. A particularly helpful way of quantitative representation of the results is the rocking curve topography \cite{Lubbert00} where a sequence of images taken at various angles on the rocking curve of the studied crystal is sorted for each pixel to obtain local rocking curves and map their parameters across regions of interest. 

\begin{figure}
\centering\includegraphics[width=0.48\textwidth]{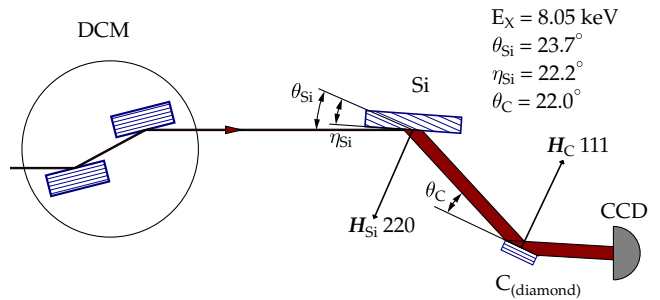}
\caption{Experimental setup for rocking curve topography (see text for details).}
\label{fig:setup}
\end{figure}

Rocking curve topography was performed at 1-BM beamline at a photon energy of $E_X \simeq$~8.05~keV. The scheme of the experimental setup is shown in Fig.~\ref{fig:setup} \cite{Stoupin_AIPP16_2}. A double-crystal Si 111 monochromator (DCM) was used for preliminary monochromatization of synchrotron radiation. Diamond (111) crystal was placed in the nearly-nondispersive arrangement with a strongly asymmetric Si (220) beam conditioner crystal (asymmetry angle  $\eta_{Si}$~=~22.2~deg).  

A special care was taken to align the reciprocal vectors of the beam conditioner and the studied diamond to the scattering plane of the experiment (vertical). This helped to minimize aberrations which reveal themselves as a horizontal gradient in the rocking curve topographs. The choice of the vertical plane corresponds to $\sigma$-polarization of the synchrotron radiation with respect to the scattering plane. After alignment of the reciprocal vectors a sequence of images of the exit beam was collected using a digital area detector (CCD) with $13 \times 13 \mu$m$^{2}$ spatial resolution. Each image in the sequence corresponded to a different angle ($\theta_C$) on the rocking curve of the diamond crystal (selected images are presented in the Supplementary section).

\begin{figure}
\centering\includegraphics[width=0.49\textwidth]{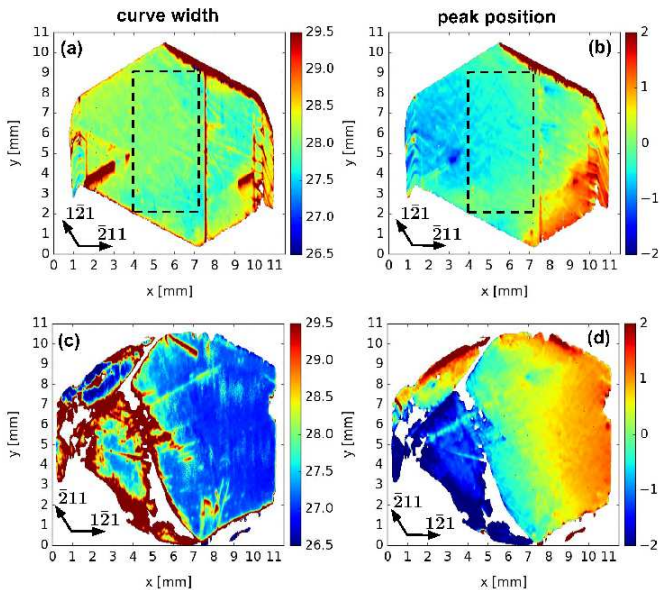}
\caption{Rocking curve topographs of plate B obtained from the as-grown (111) surface: (a) map of the curve width (FWHM) and (b) map of the curve peak position (midpoint).  The dashed box with size of $\simeq$~3~$\times~7$~mm$^2$ indicates a nearly perfect crystal region where r.m.s. variations in the curve width and the peak position are less than one microradian. Rocking curve topographs of plate B obtained from the polished side: (c) map of the curve width (FWHM) and (d) map of the curve peak position (midpoint). The units on the colorbars are microradians. The arrows in each subfigure indicate directions in the reciprocal lattice. The horizontal direction was aligned with the axis of rotation in the rocking curve measurements. The maps are scaled in the vertical direction to represent projections on the studied crystal surfaces.}
\label{fig:rctopo}
\end{figure}

Rocking curve topographs representing maps of rocking curve width and peak position were calculated using DTXRD code~\cite{dtxrd}. 
The peak position was calculated as an average of the angular positions of the negative and positive slopes of the rocking curve while the curve width was calculated as their difference. The positions of the slopes were defined at the half maximum of intensity. Figure~\ref{fig:rctopo} shows rocking curve topographs of the selected crystal plate where (a) and (b) correspond to the as-grown (111) surface and (c) and (d) correspond to the laser sliced polished surface. 

The topographs of the as-grown side reveal a region of high crystal quality (working region) with a size of $\simeq$~3~$\times~7$~mm$^2$ (dashed box in Fig.~\ref{fig:rctopo}). The root-mean-square (r.m.s.) variation in the width of the local rocking curves was found to be $\Delta \theta \simeq$~0.26~$\mu$rad while the r.m.s. variation in the rocking curve peak position was about the same $\delta \theta \simeq$~0.25~$\mu$rad. The close-up topographs of the working region are given in the Supplementary section. The measured width ($\approx$~28~$\mu$rad) and the shape of the local (per-pixel) and total rocking curves (averaged across the working region) were found to be close to theoretical calculations in the present multi-crystal geometry (see Supplementary section for details).  
Outside of the working region the stacking faults found on white-beam topographs are also present on the rocking curve topographs. 
On the rocking curve topographs taken from the polished side of the plate the faceted structure is clearly seen. The map of the curve width (Fig.~\ref{fig:rctopo}(c)) indicates presence of lattice damage at the boundaries of the facets while the map of the peak position (Fig.~\ref{fig:rctopo}(d)) suggests that the the crystal lattice is tilted at a distinct angle for each facet. In spite of these defects, the largest facet also contains a region of high crystal quality of considerable size.

\section{Discussion and Summary}
In white beam Laue X-ray topography the entire thickness of the crystal is probed with X-rays. In our characterization experiment this was confirmed  by the observation of plastic deformation (the facet structure) on the polished (back) surface of the crystal while the as-grown (front) surface was illuminated with the white beam.  
On the contrary, in the Bragg case of X-ray diffraction employed in the rocking curve topography the penetration of the X-ray wavefield into the crystal is limited by the extinction depth at the peak of the reflection rocking curve and by the attenuation depth at the tails of the rocking curve. For diamond 111 reflection at $E_X \approx$~8~keV the extinction depth is about 1 $\mu$m while the attenuation depth is $\simeq$~250~$\mu$m. 

Each of the rocking curve topographs combines images taken at different angles on the rocking curve, thus representing combined information for depths up to the attenuation depth. The images taken at the slopes of the rocking curve in the pseudo plane wave configuration test the response of the crystal planes to off-energy components of a polychromatic collimated X-ray beam of an undulator-based X-ray source. Thus, the angular mapping in rocking curve topography provides quantitative characterization of the wavefront of the reflected radiation. Since the probing depth is limited by the attenuation length, for the 1-mm-thick plate we gain insights on the diamond quality and applicability as a wavefront-preserving monochromator for each surface independently. 

The observed small angular variation $\delta\theta \simeq$~0.25~$\mu$rad across the $\simeq$~3~$\times~7$~mm$^2$ region enters a simple wavefront preservation criterion:
\begin{equation}
\delta \theta \ll \Delta \psi
\label{eq:crit}
\end{equation}
where $\Delta \psi$ is the angular divergence of the incident X-ray beam generated by the light source. For an undulator beamline of a third-generation synchrotron 
$\Delta \psi \simeq$~5~$\mu$rad r.m.s. and the criterion is satisfied. For XFEL such as the Linac Coherent Light Source (LCLS) $\Delta \psi \simeq$~1~$\mu$rad r.m.s. The criterion is still reasonably satisfied, however, further improvement is desirable. The fine structure in the working region in Fig.~\ref{fig:rctopo}(a,b) (streaks with less than 1~$\mu$rad variation) is likely related to plastic deformation on the as-grown diamond surface. An optimized diamond polishing procedure (with an introduction of a 1-2~deg~miscut angle to avoid polishing along the (111) plane) will likely yield better results. Also, improvement of angular sensitivity of the pseudo plane wave setup (with a reasonable goal of ~0.1~$\mu$rad) will be necessary, which requires a dedicated low-vibration setup. 
Finally, considering the applicability of the studied plate B for wavefront-preserving monochromators at synchrotrons it should be noted that the heat load due to absorption of incident radiation is considerably greater (by about two orders of magnitude) due to the high repetition rate (few MHz) of X-ray pulses compared to  that of the presently operational XFELs ($\approx$~100 Hz). The contribution of radiation heat-load-induced distortions of the crystal lattice  often represent a comparable or even a dominant contribution to the wavefront distortion of the reflected X-rays. A simple diagnostics method for quantitative characterization of these distortions across the beam footprint was recently demonstrated \cite{Stoupin15}. 
To minimize the heat-load-induced lattice distortion development of an efficient thermal coupling of the diamond crystal to a cooled crystal holder is required. A widely accepted solution involves the use of InGa alloy (viscous liquid metal at room temperature) which immobilizes the diamond crystal in the holder without substantial mounting-induced crystal strain while filling the gaps on contact surfaces to enhance thermal contact (e.g., \cite{Schildkamp02}). The efficiency of this coupling is clearly affected by the available surface contact area. The back surface of the working region cannot be used for the heat sink since substantial portion of the incident beam (e.g., high radiation harmonics) is transmitted through the crystal. For plate B, the working region of high quality is nearly centered on the crystal. Thus, two large portions ($\simeq$~3~$\times$~6 mm$^2$) adjacent to the crystal edges on each side from the working region can be used as thermal contact surfaces to enable an efficient heat sink.

In summary, we have shown that very large (10~$\times$~10 mm$^2$) HPHT single crystal diamond plates with (111) surface orientation are feasible. These plates were fabricated entirely from the (111) diamond growth sector. A region of high crystal quality with a size of about 3~$\times$~7~mm$^2$ was found for a selected plate. The misorientation of Bragg planes across this region was found to be less than 1 microradian. The region was nearly centered on the diamond plate which permits the use of large portions ($\simeq$~3~$\times$~6 mm$^2$) adjacent to the edges of the plate for efficient thermal coupling to a cooled crystal holder. These observations indicate that the plate is suitable for applications in high-heat-load wavefront-preserving crystal optics for modern hard X-ray light source facilities. 

\section{Acknowledgements}
K. Lang, R. Woods and J. Kirchman are acknowledged for technical support of the X-ray topography experiments. 
Use of the Advanced Photon Source was supported by the U. S. Department of Energy, Office of Science, under Contract No. DE-AC02-06CH11357.  


\end{document}